# FULL MESH NETWORKING TECHNOLOGY WITH PEER TO PEER GRID TOPOLOGY BASED ON VARIABLE PARAMETER FULL DIMENSIONAL SPACE

Wenqiang Song[1], Chuan He[2], Zhaoyang Xie[3] and Yuanyuan Chai[4]

[1]JIT Research Insititute, Beijing, China
wenqiang_song@jit.com.cn
[2]JIT Research Insititute, Beijing, China
chuan_he@jit.com.cn
[3]Beijing Insititute of Technology, Beijing, China
zhaoyangxie@bit.edu.cn
[4]JIT Research Insititute, Beijing, China
yuanyuan_chai@jit.com.cn

## ABSTRACT

*The continuous development of computer network technology has accelerated the pace of informatization, and at the same time, network security issues are becoming increasingly prominent. Networking technology with different network topologies is one of the important means to solve network security problems. The security of VPN is based on the division of geographical boundaries, but the granularity is relatively coarse, which is difficult to cope with the dynamic changes of the security situation. Zero trust network solves the VPN problem through peer to peer authorization and continuous verification, but most of the solutions use a central proxy device, resulting in the central node becoming the bottleneck of the network. This paper put forward the hard-Nat traversal formula based on the birthday paradox, which solves the long-standing problem of hard NAT traversal. A full mesh networking mechanism with variable parameter full-dimensional spatial peer-to-peer grid topology was proposed, which covers all types of networking schemes and achieve peer-2-peer resource interconnection on both methodological and engineering level.*

## 1. INTRODUCTION

Network security is an important branch of IT industry, with the goal of protecting network systems, data, and services from unauthorized access and attack[1, 2]. With the spread of the internet and the acceleration of digitalization, the importance of network security is becoming increasingly prominent. In the early days, network security mainly focused on preventing the intrusion of malicious software (such as viruses and worms[3]). However, as the means of network attacks have become increasingly complex, the scope of network security has expanded to include preventing data breach, protecting user privacy, and preventing identity theft, among other aspects.
A Virtual Private Network (VPN) is a network security technology that creates encrypted network connections, allowing users to securely access remote or public networks. The advent of VPNs can be traced back to the 1990s[4] when businesses began seeking a solution to connect remote offices and employees securely and economically. VPNs create a virtual private network on a public network (such as the internet), allowing data to be transmitted in an encrypted tunnel, thereby ensuring the security and privacy of the data.

Zero Trust is a network security model whose core concept is "never trust, always verify". The emergence of this model is a reflection on the traditional "firewall" security model. In the traditional model, companies usually set up firewalls at the boundaries of their networks, and once users pass the firewall, they can access all resources within the network. However, this model overlooks internal threats and attackers who have already compromise the network. The Zero Trust model proposes that every access request should be verified, regardless of whether users are inside or outside the network and regardless of their identity. The proposition and implementation of this model has significant implications for enhancing network security.

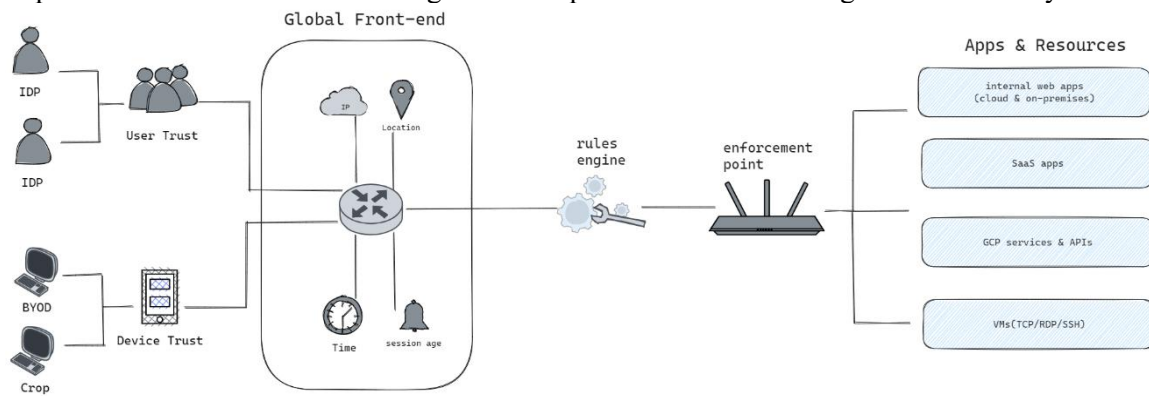

Figure 1.  Google BeyondCorp zero trust architecture

Figure 1 shows the most typical Google BeyondCorp Zero Trust topology[5]. Currently, the vast majority of Zero Trust solutions in the industry adopt this architecture or its variants. The advantage of this architecture is that it can realize all the concepts of Zero Trust. However, it highly depends on Google's enormous capability of their cloud infrastructure and assumes that all services are web-based, while not all companies meet these conditions. Therefore, we proposed a Full Mesh networking solution to implement the principles of Zero Trust without these constraints.

Since the vast majority of network nodes such as PCs, mobile devices, etc., operate most of the time in an environment without public IP, that is, behind NAT routers and firewalls, to achieve Full Mesh networking, the NAT traversal problem must be solved first.

The goal of networking lies in connecting various resources (such as people, machines, data, etc.), enabling them to communicate and share information with each other, thereby improving work efficiency and collaborative capabilities. Through networking, various functions such as remote access, remote control, file sharing, video conferencing, etc., can be achieved, meeting communication needs in different scenarios. The purpose of network security is to improve the security and reliability of the network, protect internal resources from unauthorized access and malicious attacks, and ensure the stability and availability of the network. Therefore, networking and network security technologies have a very important significance in modern information society, and are one of the important means to promote the development of informatization and promote economic and social progress.

VPN and Zero Trust networking[6] are the two existing networking modes, each with its own characteristics. The security of VPN is based on geographical boundary, but the granularity is relatively coarse, making it difficult to cope with dynamic changes in the security situation. Zero Trust networks solve the problem of VPN through end-to-end authorization and continuous verification, but most solutions adopt centralized proxy devices, making the central node a bottleneck and single point of failure in the network. Another possible implementation is peer-to-peer full mesh communication, but it is necessary to solve the NAT traversal problem.

This paper aimed to solve the core problem in Zero Trust networking. And as a prerequisite for the implementing full mesh networking, a hard NAT traversal formula based on the birthday paradox was put forward, which solves the long-standing hard NAT traversal problem[7]. In addition, the full mesh networking technology based on variable parameter full dimensional spatial peer-to-peer grid topology proposed in this article can also solve the problems and drawbacks of zero trust networking, achieve peer-to-peer resource interconnection, and meet the network communication requirements of full mesh networking, covering all types of networking solutions such as site to site networking.

## 2. NETWORKING REQUIREMENTS

Network Address Translation (NAT) is an address translation technology that can modify the IP address in the header of an IP datagram to another IP address, and achieve address reuse by using the translated port number. NAT is widely used as a transitional technology to alleviate the exhaustion of IPv4 public network addresses, due to its simple implementation. However, NAT also poses a potential security risk, as it can make it difficult to trace the origin of network traffic and can be used to hide malicious activities. Therefore, it is important to implement appropriate security measures, such as firewalls and intrusion detection systems, to ensure the security of networks that use NAT.

### 2.1. VPN networking approach

A VPN (Virtual Private Network) is typically used to securely connect networks between different locations, allowing access to resources in a private network from a remote location. Common use cases include employees working from home, establishing secure connections between companies and partners to share resources, and protecting network traffic from eavesdropping when using public Wi-Fi networks. To establish a VPN connection, a VPN server, which is typically located within the private network, and a VPN client, which can run on personal computers, mobile phones, or tablets, are required.

### 2.2. Zero trust networking approach

Zero Trust Network (ZTN) is a network architecture designed to protect network resources by implementing strict identity verification and access control. In the ZTN architecture, any user or device must pass multiple layers of identity verification and security checks before accessing network resources. Compared to traditional VPN network architecture[8], the advantage of ZTN is that it can effectively prevent malicious attackers from using authorized users or devices to launch attacks. For example, even if a malicious attacker obtains the account credentials of an authorized user, they still cannot access network resources because they must pass other security checks to obtain access. The application scenarios of ZTN typically include enterprise internal networks that protect sensitive data, multi-tenant environments of public cloud service providers, and networks of government and military organizations.

### 2.3. The difference between zero trust networking and VPN networking

Zero Trust and VPN are both technologies used to establish secure connections between two computers. However, they have some significant differences:
1. Zero Trust is a cloud-based architecture that allows data exchange between different organizations without a common trust basis. In contrast, VPN is a technology used to establish a secure network connection between two organizations, usually assuming some form of trust relationship between them.
2. Zero Trust typically uses encryption to protect the privacy and integrity of data, and uses authentication and authorization techniques to ensure that only authorized users can access data. VPN also uses encryption to protect data, but it also uses Virtual Private Network (VPN) protocols to hide users' internet activity.
3. Zero Trust architecture is typically used to share data between different organizations, such as in healthcare, financial services, or government agencies. VPN is typically used to connect remote users to enterprise networks or to connect two enterprise networks together.
4. Zero Trust architecture typically requires specialized software or hardware to implement, while VPN can be implemented using software or hardware or through third-party services.

## 2.4. Issues with existing networking methods

The security of VPN is based on the division of geographical boundaries (intranet and internet), which has a relatively coarse granularity. Once inside the VPN boundary, access to the entire system is allowed. The security authentication of VPN is static and cannot respond well to the dynamic changes in security situations[9].

Zero Trust solves the problems of VPN by implementing end-to-end authorization and continuous verification. However, most Zero Trust solutions typically use a centralized proxy device to proxy traffic to access services. Although this solves the inherent problems of VPN's boundary division and continuous verification, the centralized topology of the proxy device causes it to become a bottleneck and a single point of failure in the network. Another possible implementation of Zero Trust[10] is for all communication nodes to implement point-to-point full mesh communication with each other, which can overcome the problems of VPN and avoid the typical issues of centralized topology in Zero Trust solutions. However, due to the existence of a large number of NAT devices in the current network, the problem of NAT traversal needs to be solved first to achieve truly feasible full mesh communication.

**Contribution：**
1. NAT traversal has been a long-standing problem in the industry without a perfect solution. This article proposes a hard-NAT traversal formula based on the birthday paradox.
2. This article proposes an implementation solution for SDP (Software Defined Perimeter) based on the birthday paradox theory, named full mesh networking based on the variable parameter full dimensional spatial peer-to-peer grid topology, which meets the requirements of full mesh networking.

In this section, we introduced the networking requirements and two different networking methods: VPN and Zero Trust networking. Zero Trust networking solves the problems of VPN by implementing end-to-end authorization and continuous verification, but it has the bottleneck and single point of failure problem[] caused by centralized proxy devices. Therefore, we propose a hard-NAT traversal formula based on the birthday paradox and a full mesh networking SDP solution to meet all the requirements of networking.

## 3. THE HARD-NAT TRAVERSAL PROBLEM AND FORMULA

### 3.1. The hard-nat problem

The traversal problem occurs when two private networks want to communicate over the Internet, and the NAT device is unable to properly route the packets to the correct destination because they are both using private IP addresses. The most common scenario for this problem is when both devices are on different private networks and they cannot communicate directly because their private IP addresses cannot be properly forwarded to each other over the Internet.

There are two types of traversal problems: soft NAT traversal and hard NAT traversal. Soft NAT traversal is usually caused by a NAT device that is not properly configured or does not have UPnP turned on.UPnP (Universal Plug and Play) is a universal network protocol that allows devices to automatically configure port mapping rules so that ports can be opened and closed automatically when needed. If a NAT device does not have UPnP enabled or does not configure the port mapping rules correctly, this can lead to soft NAT traversal problems.

The hard NAT refers to a stricter form of Network Address Translation (NAT), also known as Symmetric NAT. In hard NAT, the NAT device assigns each connection a unique port number that can only be used for that connection and cannot be used by any other connection. This assignment results in external devices not being able to directly access devices in the private network, which can lead to hard NAT traversal problem. By using asymmetric port mapping, hard NAT makes it impossible for external devices to directly access devices on the private network. When a device on a private network wants to communicate with an external device, it usually needs to use some special techniques and protocols, such as STUN, TURN, ICE, etc., to solve the hard NAT traversal problem.

In easy NAT, the NAT device assigns each internal device a public IP address and port number that is unique to that device, and external devices can access that device through that address and port number. Compared to hard NAT, easy NAT uses a relatively loose port mapping method, which makes it easier for external devices to access devices on the private network. When a device initiates a connection to the outside, the NAT device uses the public IP address and port number of this device to map this connection, and when an external device initiates a connection to this device, the NAT device decides which device to forward this connection to based on the destination IP address and port number of the connection.easy NAT is a relatively loose NAT translation method that uses a relatively loose port mapping method, which makes it relatively easy for external devices to access devices on the private network. However, easy NAT also has some security issues, and the appropriate security configuration should be considered.

We call hard NAT and its variants "Endpoint-Dependent Mapping" (EDM). But hard NAT is a big problem for us, as long as there is such a device in the path, the previous scheme will not work. In addition, certain networks block NAT traversal, which has a much greater impact than this hard NAT. For example, we found that UCBerkeleyguestWiFi blocks all outgoing UDP traffic except DNS traffic. No matter what NAT hacks are used, there is no way to get around this block. So we need a reliable fallback mechanism.

This section discusses the NAT traversal problem in the network, including soft NAT traversal and hard NAT traversal, and two types of NAT translation methods, easy NAT and hard NAT. For the hard NAT traversal problem, the use of techniques and protocols such as STUN, TURN, and ICE are proposed to solve the problem. However, some networks that block NAT traversal would require a reliable fallback mechanism.

### 3.2. The hard-NAT traversal formula based on the birthday paradox

The main problem is that the easy NAT side does not know which address (IP port combination) to send data to on the hard NAT side, but must also send data to the hard NAT side to open the firewall on that side.

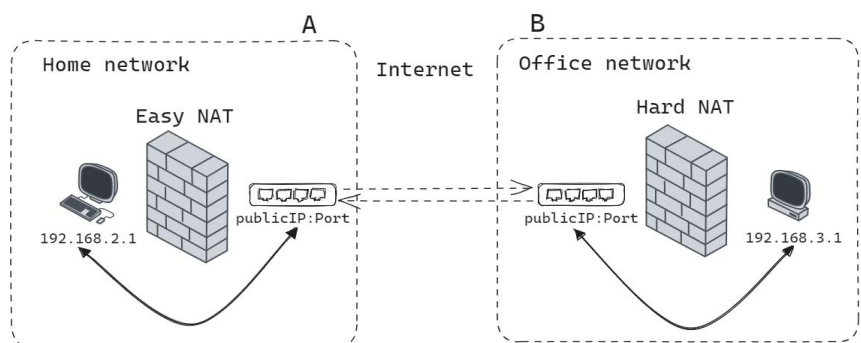

Figure 2. easy NAT and hard NAT traversal

Figure 2 shows that a specific IP and port on this side of the hard NAT because it has been pre-stunned. Assuming for a moment that the IP address is correct, then it is the port that needs to be addressed. There are 65535 possible port numbers. We can scan them one by one and find the correct port number in 10 minutes at worst, if we scan 100 per second. It can solve the problem, but not very cleverly. And it looks so much like port scanning to the IDS software (because that's what we're actually doing) that it's basically going to be blocked.

Using the birthday paradox theory, we can do much better than port scanning! Instead of scanning 65535 possible ports one by one, we can open 256 ports at once on the hard NAT side by establishing 256 sockets which can send data to the easy NAT side and let the easy NAT side randomly probe the target ports.

The birthday paradox is the probability that at least two people out of no less than 23 people have the same birthday is greater than 50%. For example, in an elementary school class of 30 students, the probability of two people having the same birthday is 70%. In a large class of 60 students, the probability is greater than 99%. The birthday paradox is a "trick" in the sense that it creates a logical contradiction. However, this mathematical fact is so counterintuitive that it is called a paradox. The mathematical theory of the birthday paradox has been applied to the design of a cryptographic attack method - the birthday attack.

In the hard NAT traversal problem, A side is easy NAT and B side is hard NAT, the ports of A are fixed (one and known), and B hypothetically opens 256 ports (but it is impossible to know what these 256 port numbers are), which we can scan a total of m (m=t*R) times.t is the scan time and R is the scan frequency.

For example, if A side opens the port 18888 (which is fixed and known) and B side opens 256 ports ( with the port numbers unknown), then a scan (assuming port B is port 1025) can simply be viewed as an attempt to connect on port A:18888 -> B:1025 to see if the number 1025 exists among the 256 ports opened on the B side.

So if we consider the total number of ports that can be used in the end to be from 1025 to 65535, then the problem can be simplified as following: there are a total of (65535-1024) balls in a pool, of which there are B black balls, and the probability that we will catch the black ball if we catch it A times is the results we want. Here B is the number of ports opened on the B side, for example 256, A is the number of times the A side probed.Based on the birthday paradox, the hard NAT traversal formula (1) is as following:

$$P = 1 - \prod_{i=0}^{A-1} \frac{K - B - i}{K - i} \qquad (1)$$

Where, P is the final calculated probability that it can be successfully traversed, the constant K is the total number of available ports (from 1025 to 65535), A is the number of probes on the A side (i.e., scan time *scan frequency: t*R), and B is the number of open ports on the B side (e.g., 256).

Table 1 shows the probability of success based on the classic birthday paradox (assuming 256 ports opened on the hard NAT side and at a rate of 100 times/s). according to the formula (1) proposed in this paper, when the number of ports is fixed at 256, the success probability can be obtained as more than 98% when the number of subsequent probes is 1000, i.e. more than that. In practical application, the above conclusion can also be obtained. Therefore, this traversal formula provides a theoretical basis for the proposed full mesh networking scheme.

Table 1. Heading and text fonts.

| Cost time of probes | Probe times | Success probability | Failure probability |
| --- | --- | --- | --- |
| 5s | 500 | 86.41018% | 13.58982% |
| 10s | 1000 | 98.18191% | 1.81809% |
| 15s | 1500 | 99.76061% | 0.23939% |
| 20s | 2000 | 99.96899% | 0.03101% |

Figure 3 shows the variation of connection success probability with the number of random probes for 128, 256, and 512 ports opened in hard nat. Figure 3 compares the number of probes required to achieve 99% success probability when different numbers of ports are opened in hard nat. Notice that the higher the number of opened ports, the less probes are needed to reach 99% success probability. Based on engineering experience and resource consumption in real-world usage, we generally use 256 as the number of opened ports in the hard side for NAT traversal.

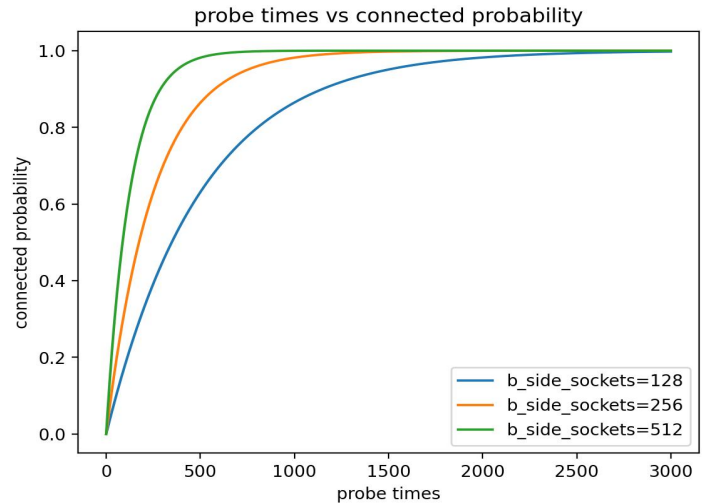

Figure 3. the probability of a successful connection as a function of the number of random probes (with 128, 256, and 512 ports opened in hard nat, respectively).

This section describes how to use the birthday paradox theory to solve the difficult problem of determining the destination port when communicating data between easy NAT and hard NAT. By opening multiple ports at the same time, the birthday paradox theory is used to gradually narrow down the range of possible ports and finally determine the target port. The traversal formula proposed in this article can calculate the probability of successful traversal. When the number of random detections is over 1000, the probability of success can reach over 98%, providing a theoretical basis for implementing the networking scheme in the next section.

## 4. FULL MESH NETWORKING TECHNOLOGY BASED ON VARIABLE PARAMETER FULL DIMENSIONAL SPACE

This chapter mainly discusses the full mesh networking scheme based on variable parameter full dimensional space that can be realized by using the birthday paradox based NAT traversal technology on the basis of NAT traversal capability, gateway requirements and encryption requirements. These networking schemes basically cover all existing VPN network application scenarios and have high flexibility and scalability.

## 4.1. Preliminaries

The NAT traversal formula (2) based on birthday paradox can be simplified as following :

$$P = det(t, R, n) \qquad (2)$$

Where, P represents the total traversal rate, t represents the total scanning time, R represents the scanning rate (times/s), and n represents the total number of ports scanned. Usually, n is taken as 256. According to the previous conclusion, under the condition of limiting R to 100 times per second, the P value can reach 50% within 2 seconds of t, and P value can be above 99.9% before t reaches 20 seconds.

Based on the calculation of NAT traversal capability, which is P, the full mesh networking scheme based on variable parameter full dimensional space can be summarized as formula (3):

$$T = hom(G, P, \theta) \qquad (3)$$

Where, G is Gateway, in which 0 means a network without gateway and 1 means a network with a gateway. P is the NAT traversal rate, in which 0 means unsuccessful traversal and 1 means successful traversal. θ is end-2-end encryption, in which 0 means end-2-end encryption is not in place and 1 means end-2-end encryption is in place.

## 4.2. Full Mesh Networking Scheme

The full mesh networking technology based on variable parameter full dimensional space proposed in this article can comprehensively cover the following four networking schemes at both the theoretical level and engineering level. The details are shown below.

### 4.2.1. Point-2-Site

According to formula (3), when G=1 and P=0, θ=1, a Point-2 Site networking scheme can be formed.

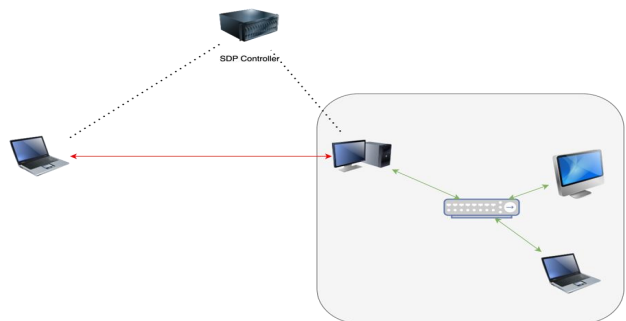

Figure 4. Figure of Point-2-Site

As Figure 4 show, on the basis of the full mesh networking, if the nodes in a local area network cannot access the trusted network through the SDP Agent totally, then access can still be achieved by using the SDP Agent as a subnet agent. The cost is that the communication between nodes in the LAN and the SDP Agent is not protected by encryption.

#### 4.2.2. Site-2-Site

According to formula (3), when G=1 and P=0, θ= 0, a Site-2-Site scheme can be formed.

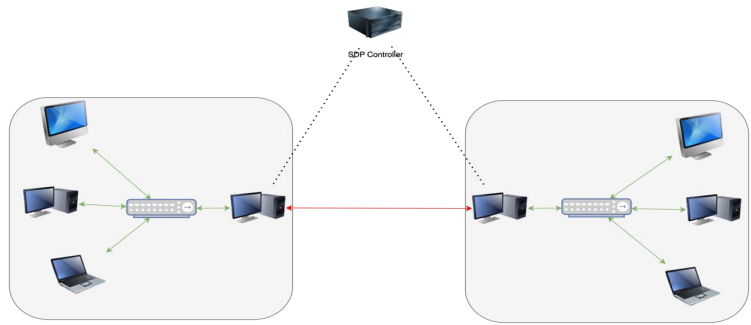

Figure 5.  Figure of Site-2-Site

As Figure 5 show, this topology is a typical VPN connection topology, and full mesh SDP can be achieved through subnet forwarding capability. If there are multiple local area networks that need to be connected, the site mesh scheme in the next section can be a reference.

#### 4.2.3. Site Mesh

According to formula (3), when G=1 and P=1, θ= 1, a Site Mesh scheme can be formed.

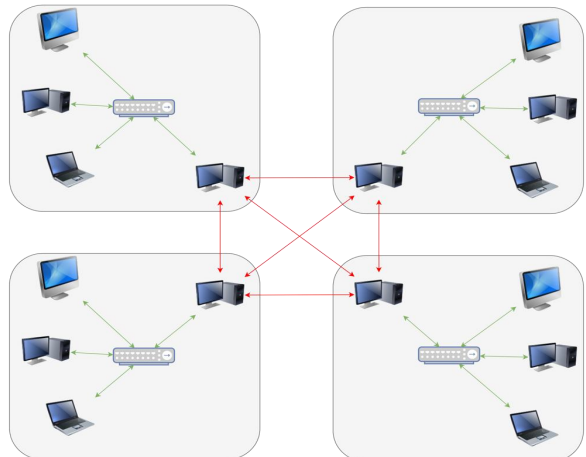

Figure 6.  Figure of Site Mesh

As Figure 6 show, on the basis of the site-2 site scheme, the forwarding ability of the full mesh SDP agent can be utilized to achieve mesh interconnection between multiple sites. The site mesh capability surpasses the ability of ordinary VPN networking, allowing multiple sites to achieve their own interconnection.

#### 4.2.4. Full Mesh

According to formula (3), when G=0 and P=1, θ= 1, a Full Mesh scheme can be formed.

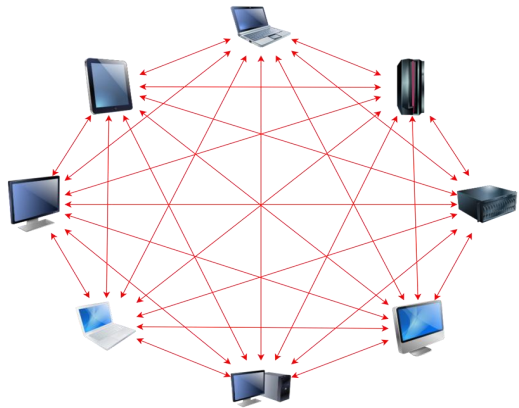
Figure 7. Figure of Full Mesh

Full mesh scheme is the most ideal network form that meets all zero trust requirements, as shown in the figure 7. Each computing node (including physical and virtual) joins a peer-to-peer fully connected network through an SDP agent, and the connection between any two points is encrypted and access permissions are individually separately.

Full mesh networking has the following advantages:
1. High reliability: Due to each device in the network being directly connected to each other device, there are multiple data transmission paths. This means that if a device or link fails, there are still other paths available for data transmission, making the network highly reliable.
2. High bandwidth: Networks can have high bandwidth because each device is directly connected to each other device. This means that there is no need to route data through intermediate devices, which can save time and improve overall performance.
3. Easy setting up: The complex routing protocols or network infrastructure are not reqired. All the required content is directly connected to each device.
4. Scalability: it is easy to add new devices to the network. All the required content is a direct connection between the new device and every other device in the network.
5. Flexibility: Networks are very flexible because they can be easily modified to adapt to changes in the network. For example, if a device needs to be removed from the network, simply disconnect it, and other devices in the network will continue to operate unaffected.

## 5. CONCLUSION

We first discussed networking requirements and two different types of network configurations: VPN and Zero Trust networking. Zero Trust networking solves the issues of VPN through end-to-end authorization and continuous verification, but it presents problems of bottlenecks and single point failures due to centralized proxy devices. Therefore, we have proposed a Hard-NAT traversal formula based on the birthday paradox and Mirage SDP solutions that can meet the requirements of Full Mesh networking.

When discussing the NAT traversal issue, we introduced the concepts of Soft-NAT and Hard-NAT traversal and compared the two NAT traversal methods, Easy-NAT and Hard-NAT. For the Hard-NAT traversal problem, we suggested using technologies and protocols such as STUN, TURN, ICE, and also proposed a fallback mechanism to cope with situations where some networks may block NAT traversal entirely.

Next, we detailed the network penetration technology based on the birthday paradox, which can solve the problem of being unable to determine the target port when data communication occurs between easy NAT and hard NAT. By opening multiple ports simultaneously, we use the theory of the birthday paradox to gradually narrow down the possible range of ports, ultimately determining the target port. The formula we proposed can calculate the probability of successful penetration, with a success rate of over 98% when the number of random probes exceeds 1000, providing a theoretical principle for the networking scheme to be implemented in the next section. Finally, we discussed the variable parameter full-dimensional peer-to-peer networking schemes that can be achieved using the network penetration technology based on the birthday paradox. These networking schemes basically cover all existing network application scenarios of VPNs and have high flexibility and scalability. Through this section's introduction, readers can better understand networking requirements, the NAT traversal issue, and the network penetration technology utilizing the birthday paradox, thus better addressing actual network application scenarios.

## ACKNOWLEDGEMENTS

Thanks to the teachers and researchers of the Research on satellite communication security system Project Team of the Jilin Science and Technology Office. Thanks to everone!

## REFERENCES

[1] Lee, S.hyun. & Kim Mi Na, (2008) "This is my paper", ABC Transactions on ECE, Vol. 10, No. 5, pp120-122.
[2] Gizem, Aksahya & Ayese, Ozcan (2009) Coomunications & Networks, Network Books, ABC Publishers.
[3] Vinoth, S., et al. "Application of cloud computing in banking and e-commerce and related security threats." Materials Today: Proceedings 51 (2022): 2172-2175.
[4] Mughal, Arif Ali. "Well-Architected Wireless Network Security." Journal of Humanities and Applied Science Research 5.1 (2022): 32-42.
[5] Wu, Xuesong, et al. "Threat analysis for space information network based on network security attributes: a review." Complex & Intelligent Systems (2022): 1-40.
[6] Li, Fang. "Network Security Evaluation and Optimal Active Defense based on Attack and Defense Game Model." 2023 International Conference on Distributed Computing and Electrical Circuits and Electronics (ICDCECE). IEEE, 2023.
[7] Bansal, Bijender, et al. "Big Data Architecture for Network Security." Cyber Security and Network Security (2022): 233-267.
[8] Ghelani, Diptiben, Tan Kian Hua, and Surendra Kumar Reddy Koduru. "Cyber Security Threats, Vulnerabilities, and Security Solutions Models in Banking." Authorea Preprints (2022).
[9] Hasan, Mohammad Kamrul, et al. "A review on security threats, vulnerabilities, and counter measures of 5G enabled Internet-of-Medical-Things." IET Communications 16.5 (2022): 421-432.
[10] Pramanik, Sabyasachi, et al., eds. Cyber Security and Network Security. John Wiley & Sons, 2022.
[11] Ghelani, Diptiben. "Cyber Security in Smart Grids, Threats, and Possible Solutions." Authorea Preprints (2022).

## AUTHORS

**Wenqiang(Andy) Song** got his Master degree of Computer Science from Northeastern University majoring Network Security.He is ighly experienced Computer Science professional with a specialization in Network Security. Leveraging both academic and industrial knowledge, he's worked in key roles as a Principal Engineer and Architect for leading companies, including GE, Amazon, and Alibaba Cloud. Currently, he is driving research initiatives at JIT Research Institute as part of Changchun Jilin University Zhengyuan Information Technologies Co., Ltd.

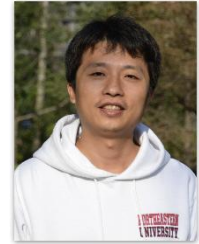

**Chuan He**, graduated from BIT in 2010 with a master's degree. He has been engaged in software development and Internet back-end service architecture for 11 years. He has worked for IBM, Sina Weibo, etc. as a senior software developer and back-end architecture, mainly engaged in mainframe system software development, advertising back-end system architecture and abtest back-end service architecture.Currently working at JIT Reserch Insititute, focusing on research and technical implementation of network security and Zero Trust Networking solutions.

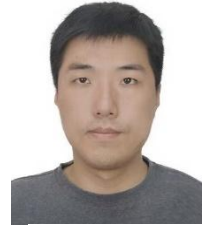

**Zhaoyang Xie** obtained his B.S. in XiDian University. He is currently a master's student at Beijing Institute of Technology and he is working as an intern at JIT Reserch Insititute. His reseach interest includes MultiParty Computing, Distributed System, Consensus Algorithms and Fully Homomorphic Encryption.

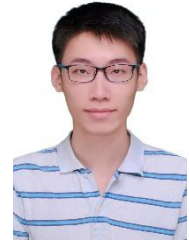

**Yuanyuan Chai** obtained her master's degree in BIT and PhD degree in Beijing Jiaotong University in 2010. She major in AI algorithm research and development and have 16 years of experience. With a profound AI theoretical research background, she have solved challenging problems with solid learning ability cross-border and rapid product conversion thinking. During 16 years of experience in IT product development and project implementation, she earned experience in R&D management, technical planning, product testing, team formation, and management. As the chief technology scholar, she also familiar with programming skill including Python, SQL, Matlab, and AI algorithm such as LR, SVM, and deep neural network model such as CNN, DNN, LSTM, Transformer, and GAN. Technology landing areas include information, traditional industry, intelligent transportation and signal communication

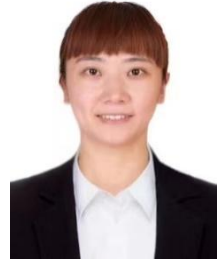